\documentclass{article}
\usepackage[textures]{graphicx}  
\begin{document}

\title{\LARGE \bf The Dynamical Velocity Superposition Effect in the Quantum-Foam In-Flow Theory of Gravity }  
\author{Reginald T. Cahill\\
{\it \small School of Chemistry, Physics and Earth Sciences}\\
 {\it \small Flinders University} \\ 
{\it \small GPO Box 2100, Adelaide 5001, Australia} \\
{\it \small(Reg.Cahill@flinders.edu.au)}}

\date{}
\maketitle

\begin{center} arXiv:physics/0407133  July 26, 2004 

To be published in {\it Relativity, Gravitation, Cosmology}\end{center}
\vspace{10mm}
\begin{abstract}
The new `quantum-foam in-flow' theory of gravity has explained numerous so-called gravitational anomalies,  
particularly the `dark matter' effect which is now seen to be a dynamical effect of space itself, and whose
strength is determined by the fine structure constant, and not by Newton's gravitational constant $G$.  Here we show 
an experimentally significant approximate dynamical effect, namely a vector superposition effect which arises
under certain dynamical conditions when  we have absolute motion and gravitational in-flows: the velocities  for
these processes are shown to be approximately vectorially additive under these conditions. This effect plays a
key role in interpreting the data from the numerous experiments that detected the absolute linear motion of the
earth.  The violations of this superposition effect lead to observable effects, such as the generation of
 turbulence.  The flow theory also leads to   vorticity effects that the Gravity Probe B gyroscope
experiment will soon begin observing. As previously reported General Relativity predicts a smaller vorticity effect
(therein called the Lense-Thirring `frame-dragging' effect) than the new theory of gravity. 
\end{abstract}

\newpage

\tableofcontents

\section{ Introduction\label{section:introduction}}

A new theory of gravity has recently been proposed \cite{NovaBook,NovaDM}, with an  earlier
zero-vorticity version given in
\cite{GQF,RGC}, that differs significantly in its effects from both the Newtonian theory of gravity and from General
Relativity, but which agrees with these theories in those very restricted cases where they have been successfully
tested, and as well gives a completely different ontology.  In the Newtonian theory gravity is explained as a
consequence of the gravitational acceleration field, while in General Relativity the explanation is in terms of the
metric of a curved `spacetime' manifold construct. In contrast, in the new theory, gravity is a consequence of a
restructuring of a substratum of space; at its deepest level space is non-geometrical and is best described as a 
processing `quantum foam'. Matter effectively dissipates this quantum-foam, and so this substratum essentially
`flows' towards matter.  In the new theory the inhomogeneities   and time-dependencies of this `flow'  manifest as the
phenomenon we know as gravity. This is a non-metric theory of gravity and, except in the special case of the external
Schwarzschild metric in General Relativity (GR), the predictions of GR and the new theory are different.  This special case 
arose in the so-called `tests' of General Relativity, such as the precession of the perihelion of planetary elliptical orbits,
the gravitational bending of light by the sun, and the gravitational redshift effect.
  
In {\it Process Physics} \cite{NovaBook, RC01} space is
essentially an `information-theoretic' system, and is a totally different category of existence from time, which is
modelled as a `process', and not as a geometrical entity.  Because of this the new `process physics' predicted that
absolute motion should have been observed
\cite{RC01}.  Absolute motion is motion relative to space itself.  A subsequent review of the experimental data
showed that indeed absolute motion had been observed  at least seven times, including even the original
Michelson-Morley experiment of 1887.  It was only in 2002 \cite{CK} that it was discovered how the Michelson
interferometer actually operated; only in the presence of a gas in the light path can this device detect absolute
motion\footnote{In vacuum the geometrical path difference effect that Michelson had originally considered is cancelled
exactly by the Fitzgerald-Lorentz contraction effect upon the arms of the interferometer, which is of course how this effect
was proposed. However in the presence of a gas this cancellation is incomplete, and some of the early interferometer
experiments were done in air or helium, so permitting the dependence on $n$ to be recently  confirmed using this older
gas-mode interferometer data
\cite{CK,AMGE}.  So modern vacuum resonator interferometers are incapable of detecting absolute motion. However putting gas
into the resonators does enable them to detect absolute motion. One such experiment is about to be performed.}, and fortunately
several interferometer experiments were done in air while two others were done in helium gas. This is explained in detail in
\cite{AMGE}, but it needs to be emphasised that the operation of the Michelson interferometer requires that the so-called
`special relativistic' length contraction effect be taken into account.  But even then the fringe shift effect is suppressed by
the very small factor
$n^2-1$, compared to the Newtonian physics theory for the interferometer that Michelson had used in the analysis of the indeed
small but not `null' fringe shifts of the 1887 experiment; here $n$ is the refractive index of the gas.  The recent 2002
theory for the Michelson interferometer enabled a detailed re-analysis of data from five such experiments, and the
extracted velocity of absolute motion was  found to be completely consistent, and also consistent  with absolute
motion experiments done using co-axial cables\footnote{Optical fibres cannot be used, see \cite{NovaBook}  for discussion.}
\cite{AMGE}, giving the velocity  of some
$417\pm40$km/s in the direction (Right Ascension $=5.2^{hr}$, Declination$ = -67^0$)  \footnote{This is different from the
Cosmic Microwave Background (CMB) anisotropy determined velocity which is, for the solar system, $369$ km/s in the direction
 $(RA=11.20^h,Dec=-7.22^0)$. These velocities are different because they relate to different effects; the absolute velocity
vector refers to absolute motion wrt local space, and is associated with the rotation of the Milky Way, the Milky
Way gravitational inflow, and also with the  motion of the Milky Way within the local cluster etc, whereas the CMB velocity is
wrt the average `universal' spatial structure, at least that spatial section presently `visible'. }.
  
The Miller interferometer experiment \cite{Miller} of 1925/1926  was so comprehensive in its data that not only
was the orbital motion of the earth about the sun detected, and used by Miller to calibrate the interferometer\footnote{And so
avoiding the use of the incorrect `Newtonian theory' used by Michelson.}, but now in conjunction with the new theory of
gravity, which affects Miller's calibration protocol and so  leading to a re-analysis of his data, it was discovered that
experimental evidence of the gravitational `in-flow' past the earth towards the sun was present in the  data \cite{AMGE}.  

The new theory of gravity involves two constants, one being
Newton's gravitational constant $G$, which now is seen to essentially determine the rate at which matter `dissipates'
the quantum foam, and a second new dimensionless `gravitational constant', which determines the self-interaction of
the quantum foam; this is a significant dynamical effect absent  in both the Newtonian theory and  General
Relativity.  Analysis of the Greenland bore hole $g$ anomaly data  \cite{NovaDM} revealed the numerical
value of this constant to be the fine structure constant, $\alpha\approx 1/137$, to within experimental
error. Then the new self-interaction dynamics was shown to explain the so-called `dark matter' effect, which
has remained a deep mystery in physics since its discovery in the relative motion of galaxies and in spiral
galaxy rotations.  

In brief summary we now note 
some of the successes of this new theory of gravity: (i) a dynamical explanation for the equivalence principle, and 
experimental evidence in support of the Lorentzian explanation for relativistic effects, namely that length
contractions,  time dilations, mass increases, etc are real physical effects {\it caused} by absolute motion of, say,
rods and clocks,  and that both the Galilean and Lorentz transformations describe reality, but must be applied to
different representations of the data, and that after proper analysis of the data we discover that the speed of
light $c$ is {\it only} the speed with respect to the local quantum-foam system. As well Process Physics appears to
offer an explanation for the so-called `dark energy' effect \cite{NovaBook}, (ii) an explanation of the bore hole
$g$ anomaly data, (iii) a new theory of black holes in which their properties are determined by the fine structure
constant, and not by $G$; these black holes are manifestly different from the black holes of General Relativity, and
are not formed by collapsing matter as in General Relativity, (iv) an explanation for the mass of the black holes
discovered at the centre of several globular clusters, including a detailed predicted effective mass for these which
is in agreement with the observational data, (v) an explanation for the  orbital velocity anomaly or `dark matter'
effect in spiral galaxies, but absent in elliptical galaxies, (vi) an explanation for the formation of spiral
galaxies based on primordial black holes, and so explaining why quasars formed so early in the universe, as
these black holes are not formed by in-falling matter, (vii) a confirmation of gravitational waves predicted
by the new theory and now apparent in data from various detections of absolute motion.  These gravitational
waves are essentially turbulence in the quantum-foam `flow', and are very different in characteristics and
speed from those predicted by General Relativity, but so far undetected. Indeed one strong prediction of the
above analyses is that both the Newtonian theory of gravity and General Relativity are falsified. In
particular this implies that the General Relativity gravitational waves do not physically  exist, (viii) the
prediction of novel effects in Cavendish laboratory experiments designed to measure
$G$. Indeed anomalies in the  existing data are already indicating the existence of spatial self-interaction
effects, and these experiments are predicted to be able to determine the value of $\alpha$, so here is a firm prediction that
a Cavendish experiment can easily check\footnote{Even a review of different past Cavendish experiments, taking into account the
new theory of gravity in Sect.\ref{section:newtheory}, should reveal the $\alpha$-dependent dynamics, and at the same time
remove the longstanding systematic discrepancies that are evident in the data.},  (ix) the prediction that the Gravity Probe B
satellite experiment
\cite{GPB} will detect precessions of the onboard gyroscopes much larger than predicted by General Relativity, and
arising from the effects of the large velocity of absolute motion of the earth upon  the so-called
`frame-dragging'.  In the new theory the `frame-dragging' is simply a result of vorticity effects caused by, in the
main, this absolute motion, with both rotational and translational motion now playing a role. As well the prediction
is made that the GP-B experiment will also be capable of detecting, subject to sufficient precession
measurement accuracy,  the new gravitational wave phenomenon
\cite{GPBwaves}, (x)  predictions that the new gravity theory requires a re-analysis of stellar dynamics, which may
have a bearing on the solar neutrino problem, (xi)  the demonstration that the new theory of gravity gives a
comprehensive explanation for the success of the Global Positioning System (GPS)
\cite{GPS} that is totally different, at an ontological level, from that of General Relativity, and that this new explanation
may enable, through the observation of subtle effects, the GPS constellation to be used to detect the new gravitational wave
phenomena, and also to improve the GPS in being used to establish the global time standard,  (xii) an explanation
for various other not well-known gravitational anomalies  such as the Allais, Saxl and Allen, Zhou, and Shnoll effects and
other observed effects, and finally (xiii) the speed of gravity is essentially infinite, that is, that a change in
position of matter results in an instantaneous change in $g$ at distant locations, as argued for in
\cite{Flandern}. This instantaneous effect results from the ongoing non-local collapse of the quantum foam
substratum to a classical state.  This non-local effect is not to be confused with gravitational wave effects,
which propagate essentially as turbulence in the flow. 

 All of the above analyses and observed
effects imply that gravity is a much more complex phenomena than is contained in either the Newtonian or General
Relativity theories.  It is also becoming clear why the deep failure of these two theories has escaped detection
for so long, over and above the long standing scandalous ban in physics on reporting on-going experimental evidence
of absolute motion and other anomalies.  In the case of the Newtonian theory  it turns out that the solar system was
too  special to have revealed the presence of the `dark-matter' effect, and then when General Relativity was being
constructed by Hilbert and Einstein, it was forced to agree with the flawed Newtonian theory in the low speed
limit. As well in the case of General Relativity all but one of the so-called tests of this theory used the
external Schwarzschild metric, and it went unnoticed that this is completely equivalent to Newton's inverse
square law. So in these cases it was still the Newtonian theory that was being tested, together with novel effects
associated  with the geodesic equation. It was not until  the in-flow formalism was recently developed that it was
realised that the so-called `new gravity dynamics' of General Relativity (GR) had in fact never been tested, except for the
one indirect case of the decay of the binary pulsar orbits, but this effect is also in the new theory\footnote{It is
a remarkable insight into the profession of physics  when only one recent astronomical observation is the sole
evidence for a theory, but  which has attracted so much attention by theoretical physicists, mathematicians, and
philosophers. }.  As George Pugh \cite{Pugh} and Leonard Schiff realised long ago  \cite{Schiff} the unique dynamical
predictions of General Relativity could only be tested in experiments such as the current Gravity
Probe B gyroscope precession experiment, and it is predicted
\cite{GPB} that the observed precessions will be very different from the General Relativity predictions.   

It is
important to understand that the new theory of gravity is totally unconnected to General Relativity, and does not use
at all the notion of a spacetime metric; indeed the very concept of spacetime is totally rejected.  In only one
special case does it turn out that mathematically the new theory may be mapped onto the mathematical formalism of
General Relativity\footnote{GR, by construction, has no `dark matter' effect, but this effect is observationally apparent in
all situations, except outside of a spherically symmetric matter distribution, but even then causes a shift in the apparent
value of $G$ \cite{NovaBook,NovaDM}. Hence the new theory of gravity cannot be mapped onto the GR formalism. However if we
make this {\it ad hoc} re-normalisation of $G$, then in the case of, for example the GPS, the new theory can be
mapped onto GR, which explains why GR, fortuitously, was successful in the design and operation of the GPS
\cite{GPS}. }, which explains why the latter supposedly had passed certain checks, but that these circumstances are
very restricted, and occur {\it only} for the external Schwarschild metric, namely external to a spherically
symmetric matter distribution.  and then only to the extent that we can ignore the small vorticity effects
associated with the absolute motion effect, which is by construction not in General Relativity.   Overall the new
theory of gravity is totally incompatible with the formalism of General Relativity, and has a totally different
ontology.  This is evident from the above list of phenomena accounted for by the new theory; all of these effects
are manifest failures of General Relativity. Indeed it has been an almost  pathological  state of affairs through
C20 physics that rotational motion was absolute, but linear motion was only relative. This can be traced back to the
incorrect conclusion  made by Michelson and Morley  when interpreting their smaller than expected fringe shifts,
i.e. based upon Newtonian physics and also based only upon 36 rotations of the interferometer, and  despite this
error being corrected by Miller in his extensive experiments of 1925/1926, involving 20,000 rotations throughout a
year, and that the time-dependence  caused by the orbital motion of the earth about the sun was evident in Miller's
data.

Herein we explain a key dynamical effect that is apparent already in observations of absolute motion \cite{AMGE},
namely that an  approximate velocity superposition effect is applicable.  This means, for example, that in the
case of the earth with, in order of decreasing magnitude, (i) a cosmic velocity of the solar system related, it seems,
to galactic and local cluster gravitational quantum-foam flows, (ii) an in-flow of space past the earth towards the
sun, (iii) the tangential orbital velocity of the earth about the sun, and (iv) a gravitational in-flow into the
earth itself, causing its own gravitational effect,  that because of special conditions that prevail in the case of
the earth that the  velocities (i)-(iii) may be added vectorially, but that velocity (iv) may not be added vectorially
and whence leads to various already observed `gravitational anomalies'.  This is of course  totally different to what
would happen in  classical `material' fluid mechanics.  There are
exceptions to this superposition  approximation which come into play under certain conditions.  One class of conditions
occurs, for example, where the in-flow velocity becomes large due to the presence of the new gravitational attractors
or `black holes' that are predicted to occur, and which have already been detected at the centres of
globular clusters \cite{NovaDM}. Another class of conditions involve 
 vorticity effects, and   the Gravity Probe B gyroscope experiment will soon begin observing spin precessions caused by
these vorticity effects. As previously reported General Relativity predicts a much smaller vorticity effect (therein
called the `frame-dragging' effect) than the new theory of gravity,   because it only includes vorticity caused by
rotation of the earth.

The significance  of the dynamical vector superposition effect is that it explains why the various observations of
absolute motion are consistent with the `in-flow' theory of gravity, and indeed why the data from the absolute motion
experiment of Miller is capable of revealing the  vector component of the in-flow past the earth towards the sun,
because the vector sum changes over the yearly orbit of the earth about the sun.

\section[  The  New   Theory of Gravity ]{The  New Theory of Gravity 
\label{section:newtheory}} 
Here  we `derive' the  `in-flow' theory of gravity by re-analysing the implications of Kepler's laws for planetary motion.
The new theory   involves a  `classical' velocity field  \cite{NovaBook,RC01} and the theory  exhibits (i) the `dark matter'
effect, with strength set by the fine structure constant, (ii) effects of absolute motion of the matter with respect to the
substratum, and (iii)  vorticity effects also caused by absolute motion of the matter, whether rotational or translational.
This flow theory is a classical description of a quantum foam substructure to space \cite{NovaBook}, and the `flow' describes
the relative motion of this quantum foam with, as we now show, gravity arising from inhomogeneities and time variations in that
flow. These gravitational effects can be caused by an in-flow into matter, or even  produced purely by the
self-interaction of space itself, as happens for instance  for the new `black holes', which do not contain in-fallen 
matter.   

The Newtonian
theory was  formulated in terms of a force field, the gravitational acceleration
${\bf g}({\bf r},t)$, and was based on Kepler's laws for the observed motion of the planets within the solar
system.  As we shall see Newton's theory of gravity is not uniquely determined by Kepler's laws when
rewritten in terms of a velocity vector field, and  introducing a unique new dynamical term we
immediately obtain the `dark matter' effect, as it has been incorrectly termed. 

In the Newtonian
theory ${\bf g}({\bf r},t)$ is determined by  the matter density
$\rho({\bf r},t)$ according to
\begin{equation}\label{eqn:g1}
\nabla.{\bf g}=-4\pi G\rho.
\end{equation}
  However there is an alternative formulation \cite{NovaBook, NovaDM} in terms of
 a vector field  ${\bf v}({\bf r},t)$ determined by
\begin{equation}
\frac{\partial }{\partial t}(\nabla.{\bf v})+\nabla.(({\bf
v}.{\bf \nabla}){\bf v})=-4\pi G\rho,
\label{eqn:CG1}\end{equation}
with ${\bf g}$ now given by the Euler `fluid' acceleration
\begin{equation}{\bf g}=\displaystyle{\frac{\partial {\bf v}}{\partial
t}}+({\bf v}.{\bf \nabla}){\bf v}=\displaystyle{\frac{d{\bf v}}{dt}}.
\label{eqn:CG2}\end{equation}
Trivially this ${\bf g}$ also satisfies (\ref{eqn:g1}).  Hence (\ref{eqn:CG1})-(\ref{eqn:CG2}) are mathematically
equivalent to (\ref{eqn:g1}).  The scalar eqn.(\ref{eqn:CG1}) can only be used to determine  a zero-vorticity
flow, $\nabla\times{\bf v}={\bf 0}$, for then we may write
${\bf v}({\bf r},t)=\nabla u({\bf r},t)$, and (\ref{eqn:CG1}) becomes  
\begin{equation}
\frac{\partial  u}{\partial t}=-\frac{1}{2}(\nabla u)^2-\Phi,
\label{eqn:NGu}\end{equation}
where $\Phi$ is the Newtonian gravitational potential determined by
\begin{equation}
\nabla^2\Phi({\bf r},t)=4\pi G\rho({\bf r},t).
\label{eqn:NGPhi}\end{equation}
It is a remarkable fact that the Newtonian theory of gravity may be exactly recast in terms of this `fluid
 flow' formalism, and even more so when in Sect.\ref{section:geodesics} it is shown that the Euler
fluid acceleration in (\ref{eqn:CG2}) arises, in the non-relativistic limit, from the usual
relativistic proper-time extremisation, which also yields the generalisation of  (\ref{eqn:CG2})  to include the
Helmholtz
 term associated with vorticity, as shown in (\ref{eqn:newaccel}). 

 Eqn.(\ref{eqn:NGu}) always has  solutions, simply because if $u({\bf r},t)$ is given at time
$t$, then by integration $u({\bf r},t)$ at later times is always uniquely determined.   However, in general
solutions of  (\ref{eqn:NGu}) are necessarily time-dependent. This is because the equation $(\nabla
u)^2=-2\Phi$, required for $\frac{\partial  u}{\partial t}=0$, does not in general have solutions.
  So the flow formalism of Newtonian gravity is in general necessarily time-dependent, and then
it is the sum of the two terms in (\ref{eqn:CG2}) which together reproduce the Newtonian prediction for
${\bf g}$.  Then if according to an observer $\rho({\bf r})$ is time-independent, then in general $u({\bf r},t)$
and  ${\bf v}({\bf r},t)$ will be time-dependent, but ${\bf g}({\bf r})$ will be time-independent.  The form for ${\bf
g}$ for another observer  in uniform motion relative to this observer is discussed later. Of course that ${\bf v}({\bf
r},t)$ is time-dependent is what would be expected of any flow-like process.  Which of these two mathematically
equivalent  formalisms of gravity is physically meaningful is determined by experiment, and numerous experiments
have detected the velocity flow, and indeed also the time-dependence of that flow \cite{RGC,CK,AMGE}; this time 
dependent flow is of course the gravitational waves of the new theory. In the restricted theory, above, these waves 
do not manifest as waves in ${\bf g}$, but once the generalisations below are included, they do so.

Most significantly we shall see that
(\ref{eqn:CG1})-(\ref{eqn:CG2}) permit a generalisation that is not possible for (\ref{eqn:g1}), and which is a
dynamical explanation of the so called `dark matter' effect. Clearly   (\ref{eqn:CG1}) cannot be the complete
equation for the flow as it would only be  sufficient  for a zero-vorticity flow.  As well the flow must 
take account of relativistic effects. 

External to a spherical mass
$M$ of radius
$R$ a static velocity field solution of (\ref{eqn:CG1}) is 
\begin{equation}
{\bf v}({\bf r})=-\sqrt{\frac{2GM}{r}}\hat{\bf r},  \mbox{\ \ }r>R,
\label{eqn:vfield}\end{equation}
which gives from (\ref{eqn:CG2}) the usual inverse square law ${\bf g}$ field
\begin{equation}
{\bf g}({\bf r})=-\frac{GM}{r^2}\hat{\bf r}. \mbox{\ \ }r>R.
\label{eqn:InverseSqLaw}\end{equation} 
\index{inverse square law}
However the flow equation (\ref{eqn:CG1}) is not uniquely determined by Kepler's laws since  
\begin{equation}
\frac{\partial }{\partial t}(\nabla.{\bf v})+\nabla.(({\bf
v}.{\bf \nabla}){\bf v})+C({\bf v})=-4\pi G\rho,
\label{eqn:CG3}\end{equation}
where
\begin{equation}
C({\bf v})=\displaystyle{\frac{\alpha}{8}}((tr D)^2-tr(D^2)),
\label{eqn:Cdefn1}\end{equation} and
\begin{equation}
D_{ij}=\frac{1}{2}(\frac{\partial v_i}{\partial x_j}+\frac{\partial v_j}{\partial x_i}),
\label{eqn:Ddefn1}\end{equation}
also has the same external solution (\ref{eqn:vfield}),  because $C({\bf v})=0$ for the flow in 
(\ref{eqn:vfield}). So the presence of the  $C({\bf v})$ dynamics would not have manifested in the special case
of planets in orbit about the massive central sun.
Here $\alpha$ is a   dimensionless constant - a new gravitational constant, in addition to the usual
 Newtonian gravitational constant $G$. However inside a spherical mass we find \cite{NovaDM} that 
$C({\bf v})\neq 0$, and using the Greenland  ice shelf bore hole $g$ anomaly data  we find that
$\alpha^{-1}=139 \pm  5 $, which gives the fine structure constant $\alpha=e^2\hbar/c \approx 1/137$
to within experimental error\footnote{The occurrence of $\alpha$ does not make these flow equations  a quantum
theory.  In QED $\alpha$ plays the role of a probability measure, and so it presumably arises in the present situation
in the same manner. This relates to the deeper information-theoretic {\it process physics} theory in which an intrinsic
stochasticity limits the information content \cite{NovaBook}.}. From (\ref{eqn:CG2}) and (\ref{eqn:CG3}) we can write
\begin{equation}\label{eqn:g2}
\nabla.{\bf g}=-4\pi G\rho-4\pi G \rho_{DM},
\end{equation}
where
\begin{equation}
\rho_{DM}({\bf r},t)=\frac{\alpha}{32\pi G}( (tr D)^2-tr(D^2)),  
\label{eqn:DMdensity}\end{equation} 
which introduces an effective `matter density' representing the flow dynamics associated with the 
$C({\bf v})$ term. However the  dynamical  effect represented by   this new term  cannot be included in the
gravitational acceleration dynamics formalism of  (\ref{eqn:g1}) because it cannot be expressed in terms of the gravitational
field ${\bf g}$.  In
\cite{NovaDM} this dynamical effect is shown to be the `dark matter' effect. 

The interpretation of the vector flow field
${\bf v}$ is that it is a manifestation, at the classical level, of a quantum substratum to space; the flow is a rearrangement
of that substratum, and not a flow {\it through}  space. However  (\ref{eqn:CG3})  requires a further generalisation to include
vorticity, and also the effects of the absolute motion of matter through this substratum. To do this a  precise
definition of what is meant by the velocity field ${\bf v}({\bf r},t)$ is needed. To be specific and also to define a
possible measurement procedure
 we can choose to  use the Cosmic Microwave Background (CMB)\index{Cosmic Microwave
Background (CMB)} frame of reference for that purpose, as this is itself easy to establish. However that does not
imply that the  CMB frame is the local `quantum-foam' rest frame. Relative to the CMB frame and using the local
absolute motion detection techniques described in \cite{NovaBook,RGC,AMGE}, or more modern techniques that are
under development, ${\bf v}({\bf r},t)$ may be measured in the neighbourhood of the observer.   Then an `object' at
location ${\bf r}_0(t)$ in the CMB frame has velocity  
 ${\bf v}_0(t)=d{\bf r}_0(t)/dt$  with respect to that  frame, and then 
\begin{equation}
{\bf v}_R({\bf r}_0(t),t) ={\bf v}_0(t) - {\bf v}({\bf r}_0(t),t),
\label{eqn:CG}
\end{equation}
where ${\bf v}_0(t)$ is the velocity of an object, at ${\bf r}_0(t)$, relative to the same frame of reference that defines
the flow field, and so ${\bf v}_R$ is the velocity of that matter relative to the substratum. To take account of the
absolute velocity of matter with respect to the local quantum foam and also of vorticity effects the flow equation
(\ref{eqn:CG3}) is  generalised to a 2nd-rank tensor equation \cite{NovaBook,NovaDM,GPB}
\begin{eqnarray}
&&\frac{d D_{ij}}{dt}+ \frac{\delta_{ij}}{3}tr(D^2) + \frac{tr D}{2}
(D_{ij}-\frac{\delta_{ij}}{3}tr D)+\frac{\delta_{ij}}{3}\frac{\alpha}{8}((tr
D)^2 -tr(D^2))\nonumber \\ && +(\Omega D-D\Omega)_{ij}=-4\pi
G\rho(\frac{\delta_{ij}}{3}+\frac{v^i_{R}v^j_{R}}{2c^2}+..),\mbox{ } i,j=1,2,3. 
\label{eqn:CG4a}\end{eqnarray}
\begin{equation}\nabla \times(\nabla\times {\bf v}) =\frac{8\pi G\rho}{c^2}{\bf v}_R,
\label{eqn:CG4b}\end{equation}
\begin{equation}
\Omega_{ij}=\frac{1}{2}(\frac{\partial v_i}{\partial x_j}-\frac{\partial v_j}{\partial
x_i})=-\frac{1}{2}\epsilon_{ijk}\omega_k=-\frac{1}{2}\epsilon_{ijk}(\nabla\times {\bf v})_k,
\label{eqn:BS}\end{equation}
and the vorticity vector field is $\vec{\omega}=\nabla\times {\bf v}$.     We obtain from
(\ref{eqn:CG4b}) the Biot-Savart form  for the vorticity field
\begin{equation}
\vec{\omega}({\bf r},t)
=\frac{2G}{c^2}\int d^3 r^\prime \frac{\rho({\bf r}^\prime,t)}
{|{\bf r}-{\bf r}^\prime|^3}{\bf v}_R({\bf r}^\prime,t)\times({\bf r}-{\bf r}^\prime).
\label{eqn:omega}\end{equation} 
Then (\ref{eqn:CG4a}) becomes an integro-differential equation for the velocity field\footnote{The superscript
notation $v_R^i$ is purely for simplicity of layout.}. The form of these equations was determined by requiring
that in the non-relativistic limit they reduce to (\ref{eqn:CG3}), in which case the vorticity must go to zero.  The
form of the vorticity dynamics in (\ref{eqn:CG4b}) follows from dimensional considerations\footnote{In General
Relativity the vorticity is only generated by  absolute  rotation, and not by absolute linear motion. This peculiarity
had its  origins in the misunderstanding of the significance of the small fringe shifts seen in the
Michelson-Morley experiment. This restriction to rotational motion is claimed to be explainable by the Mach principle. 
In the new theory, herein, motion is relative to the local quantum-foam system, i.e. the local space, whether rotational
or linear. }.

 As discussed  in \cite{GPB}   (\ref{eqn:omega}) explains the so-called 
Lense-Thirring `frame - dragging'  effect in terms of this vorticity in the flow field, but makes predictions very
different from General Relativity.  These conflicting predictions will soon be tested by the Gravity Probe B
\cite{Pugh,Schiff} satellite experiment.   However the smaller component of the frame-dragging effect caused by
the earth  absolute rotation  component of ${\bf v}_R$ has been determined  from the laser-ranged satellites
LAGEOS(NASA) and LAGEOS 2(NASA-ASI) \cite{Ciufolini}, and the data implies the indicated coefficient on the RHS
of (\ref{eqn:CG4b}) to $\pm10\%$.  However that  experiment cannot detect the larger 
component of the `frame-dragging' or vorticity  induced by the absolute linear motion component of the earth's
${\bf v}_R$  as that effect is not cumulative, while the rotation induced component is
cumulative. For that reason we must wait for the GP-B data to fully confirm the RHS of (\ref{eqn:CG4b}).  Of
course in General Relativity the absolute linear motion induced vorticity is absent.

Eqns.(\ref{eqn:CG4a})-(\ref{eqn:CG4b}) only make sense if  
${\bf v}_R({\bf r},t)$ for the matter at location ${\bf r}$  is specified. We now consider the special case where
the matter is subject only to the effects of motion with respect to the quantum-foam velocity-field
inhomogeneities and  variations in time, which causes the  acceleration which we know as `gravity'.

We note that the first serious attempt to construct a `flow' theory of gravity was by Kirkwood \cite{RK1,RK2}.
However the above theory, as expressed in (\ref{eqn:CG4a})-(\ref{eqn:CG4b}), is very different
to Kirkwood's proposal\footnote{Kirkwood constructed, supposedly from `first principles', what turned out to be an
exact  velocity field representation of Newtonian gravity, and so {\it ipso facto} missed the `dark matter'
generalisation, the absolute motion effect, the relativistic terms on the RHS of
(\ref{eqn:CG4a})-(\ref{eqn:CG4b}), and also the vorticity dynamics.}. We also note that (\ref{eqn:CG4a}) and
(\ref{eqn:CG4b}) need to be further generalised  to take account of the cosmological-scale effects, namely that the
spatial system is compact and growing, as discussed in \cite{NovaBook}. The investigation of possible non-flow substratum
effects has  been considered in \cite{CM} who consider an energy-dependent metric theory.

\section{ Geodesics \label{section:geodesics}}

Process Physics \cite{NovaBook} leads to the Lorentzian interpretation\index{Lorentzian interpretation} of
so called `relativistic effects'.  This means that the speed of light \index{speed of light} is only `c' with
respect to the quantum-foam system, and that time dilation effects for clocks and length contraction effects for
rods are caused by the motion of clocks and rods relative to the quantum foam. So these effects are real dynamical
effects caused by motion through the classicalising quantum foam, and are not to be interpreted as non-dynamical
spacetime effects as suggested by Minkowski and Einstein.  

To arrive at the dynamical description of the various effects of the
quantum foam we shall introduce conjectures that essentially lead to a phenomenological description of these effects. In
the future we expect to be able to derive this dynamics directly from the Quantum Homotopic Field Theory (QHFT)
 that describes the quantum foam system \cite{NovaBook}. Here we shall conjecture that the path of an object
through an inhomogeneous and time-varying quantum-foam is determined, at a classical level,  by a variational
principle, namely that  the travel time is extremised for the physical path ${\bf r}_0(t)$, which
presumably would arise from the wave nature of the `matter'. The travel time is defined by 
\begin{equation}
\tau[{\bf r}_0]=\int dt \sqrt{1-\frac{{\bf v}_R^2}{c^2}},
\label{eqn:f4}
\end{equation}  
with ${\bf v}_R$ given by (\ref{eqn:CG}). So the trajectory will be independent of the mass of the object,
corresponding to the equivalence principle.  Under a deformation of the trajectory  $${\bf r}_0(t) \rightarrow  {\bf
r}_0(t) +\delta{\bf r}_0(t), \mbox{\ \  we have \ \ }
{\bf v}_0(t) \rightarrow  {\bf v}_0(t) +\displaystyle\frac{d\delta{\bf r}_0(t)}{dt},$$  and also
that
\begin{equation}\label{eqn:G2}
{\bf v}({\bf r}_0(t)+\delta{\bf r}_0(t),t) ={\bf v}({\bf r}_0(t),t)+(\delta{\bf
r}_0(t).{\bf \nabla}) {\bf v}({\bf r}_0(t))+... 
\end{equation}
Then
\begin{eqnarray}\label{eqn:G3}
\delta\tau&=&\tau[{\bf r}_0+\delta{\bf r}_0]-\tau[{\bf r}_0]  \nonumber\\
&=&-\int dt \:\frac{1}{c^2}{\bf v}_R. \delta{\bf v}_R\left(1-\displaystyle{\frac{{\bf
v}_R^2}{c^2}}\right)^{-1/2}+...\nonumber\\
&=&\int dt\frac{1}{c^2}\left({\bf
v}_R.(\delta{\bf r}_0.{\bf \nabla}){\bf v}-{\bf v}_R.\frac{d(\delta{\bf
r}_0)}{dt}\right)\left(1-\displaystyle{\frac{{\bf v}_R^2}{c^2}}\right)^{-1/2}+...\nonumber\\ 
&=&\int dt \frac{1}{c^2}\left(\frac{{\bf v}_R.(\delta{\bf r}_0.{\bf \nabla}){\bf v}}{ 
\sqrt{1-\displaystyle{\frac{{\bf
v}_R^2}{c^2}}}}  +\delta{\bf r}_0.\frac{d}{dt} 
\frac{{\bf v}_R}{\sqrt{1-\displaystyle{\frac{{\bf
v}_R^2}{c^2}}}}\right)+...\nonumber\\
&=&\int dt\: \frac{1}{c^2}\delta{\bf r}_0\:.\left(\frac{({\bf v}_R.{\bf \nabla}){\bf v}+{\bf v}_R\times({\bf
\nabla}\times{\bf v})}{ 
\sqrt{1-\displaystyle{\frac{{\bf
v}_R^2}{c^2}}}}  +\frac{d}{dt} 
\frac{{\bf v}_R}{\sqrt{1-\displaystyle{\frac{{\bf
v}_R^2}{c^2}}}}\right)+...
\end{eqnarray}
  Hence a 
trajectory ${\bf r}_0(t)$ determined by $\delta \tau=0$ to $O(\delta{\bf r}_0(t)^2)$ satisfies 
\begin{equation}\label{eqn:G4}
\frac{d}{dt} 
\frac{{\bf v}_R}{\sqrt{1-\displaystyle{\frac{{\bf v}_R^2}{c^2}}}}=-\frac{({\bf
v}_R.{\bf \nabla}){\bf v}+{\bf v}_R\times({\bf
\nabla}\times{\bf v})}{ 
\sqrt{1-\displaystyle{\frac{{\bf v}_R^2}{c^2}}}}.
\label{eqn:vReqn}\end{equation}
Let us now write this in a more explicit form.  This will
also allow the low speed limit to be identified.   Substituting ${\bf
v}_R(t)={\bf v}_0(t)-{\bf v}({\bf r}_0(t),t)$ and using 
\begin{equation}\label{eqn:G5}
\frac{d{\bf v}({\bf r}_0(t),t)}{dt}=\frac{\partial {\bf v}}{\partial t}+({\bf v}_0.{\bf \nabla}){\bf
v},
\end{equation}
we obtain
\begin{equation}\label{eqn:G6}
\frac{d}{dt} 
\frac{{\bf v}_0}{\sqrt{1-\displaystyle{\frac{{\bf v}_R^2}{c^2}}}}={\bf v}
\frac{d}{dt}\frac{1}{\sqrt{1-\displaystyle{\frac{{\bf v}_R^2}{c^2}}}}+\frac{\displaystyle{\frac{\partial {\bf
v}}{\partial t}}+({\bf v}.{\bf \nabla}){\bf v}+({\bf \nabla}\times{\bf v})\times{\bf v}_R}{ 
\displaystyle{\sqrt{1-\frac{{\bf v}_R^2}{c^2}}}},
\end{equation}
and  finally 
\begin{equation}\label{eqn:newaccel}
 \frac{d {\bf v}_0}{dt}=-\frac{{\bf
v}_R}{1-\displaystyle{\frac{{\bf v}_R^2}{c^2}}}
\frac{1}{2}\frac{d}{dt}\left(\frac{{\bf v}_R^2}{c^2}\right)
+\left(\displaystyle{\frac{\partial {\bf v}}{\partial t}}+({\bf v}.{\bf \nabla}){\bf
v}\right)+({\bf
\nabla}\times{\bf v})\times{\bf v}_R.
\end{equation}
This is a generalisation of the acceleration in (\ref{eqn:CG2}) to include the vorticity effect, as the last term -
also known as the Helmholtz acceleration in fluid flows, and the first term which  is the resistance to acceleration
caused by the relativistic `mass' increase effect. This term leads to the so-called geodetic effects.
  The vorticity term causes the GP-B gyroscopes to develop the vorticity induced precession
\cite{GPB}, which is simply the rotation of space carrying the gyroscope along with it, compared to
more distant space which is not involved in that rotation.   The middle
term, namely the acceleration in (\ref{eqn:CG2}), is simply the usual Newtonian gravitational acceleration,
but now seen to arise from the inhomogeneity and time-variation of the flow velocity field. As already noted it was
this geodesic equation that has been checked in various experiments, but always, except in the case of the binary
pulsar slow-down, with the velocity field given by  the Newtonian `inverse square law' equivalent form in
(\ref{eqn:vfield}). As discussed elsewhere \cite{NovaBook,NovaDM} this flow is exactly equivalent to the external
Schwarzschild metric. 

  Note that the occurrence of
$1/(1-{\bf v}_R^2/c^2)$ in (\ref{eqn:newaccel}) will lead  to horizon effects wherever  $|{\bf v}| = c$: the region
where  $|{\bf v}| < c$ is inaccessible from the region where $|{\bf v}|>c$.  Also (\ref{eqn:f4}) is easily used to
determine the  clock rate offsets in the GPS satellites\index{Global Positioning System (GPS)}, when the in-flow is
given by (\ref{eqn:vfield}). 

So the fluid flow dynamics in  (\ref{eqn:CG4a}) and (\ref{eqn:CG4b}) and the gravitational
dynamics for the matter in  (\ref{eqn:vReqn}) now form a closed system.  
This system of equations is a considerable
generalisation from that of Newtonian gravity, and is very different from the curved spacetime
metric formalism of General Relativity. 

The above may  be modified when the `object' is a massless photon, and the
corresponding result leads to the gravitational lensing effect. But not only will ordinary matter produce such
lensing, but the effective `dark matter' density will also do so, and that is relevant to the recent observation by
the weak lensing technique of the so-called `dark matter' networks. 

\section{ The Velocity Superposition Effect\label{section:thesuperposition}}

Despite being non-linear (\ref{eqn:CG4a})-(\ref{eqn:CG4b}) possess an approximate
superposition effect, which explains  why the existence of absolute motion and as well the presence of
the $C({\bf v})$ term appear to have almost escaped attention in the case of gravitational experiments  near the
earth. 

First note that  in analysing (\ref{eqn:CG4a})-(\ref{eqn:CG4b}) we need to recognise
two distinct effects: (i) the effect of a change of description of the flow when changing between 
observers, and (ii) the effects of absolute motion of the matter with respect to the quantum foam
substratum. Whether the matter is at rest or in absolute motion with respect to this substratum does
have a dynamical effect, and this paper is primarily about understanding this effect.  While the Newtonian theory and
GR both offer an account of the first effect, and different accounts at that, neither have the second dynamical
effect, as this is a unique feature of the new theory of gravity.  Let us  consider the first effect, as this is
somewhat standard. It basically comes down to noting that under a change of observer
(\ref{eqn:CG4a})-(\ref{eqn:CG4b}) transform covariantly under a Galilean transformation.
Suppose that according to one observer $O$ the matter density is specified by  a form $\rho_O({\bf r},t)$,
and that (\ref{eqn:CG4a})-(\ref{eqn:CG4b}) has a solution
${\bf v}_O({\bf r},t)$, and then  with acceleration ${\bf g}_O({\bf r},t)$ given by (\ref{eqn:G6})\footnote{Note
that here and in the following, except where indicated, the subscripts  are $O$ and not $0$.}.  Then
 for  another observer $O^\prime$ (and for  simplicity we assume that the observers  use coordinate axes that
have the same orientation, and that at time $t=0$ they coincide), moving with uniform velocity ${\bf V}$ relative
to observer $O$,  observer $O^\prime$ describes the matter density with the form
$\rho_{O^\prime}({\bf r},t)=  \rho_O({\bf r}+{\bf V}t,t)$. Then, as we now show, the corresponding  solution to
(\ref{eqn:CG4a})-(\ref{eqn:CG4b}) for $O^\prime$ is {\it exactly} 
 \begin{equation}\label{eqn:Vsum}
{\bf v}_{O^\prime}({\bf r},t)={\bf v}_O({\bf r}+{\bf V}t,t)-{\bf V}.
\end{equation}
This is easily established by substitution of (\ref{eqn:Vsum}) into
(\ref{eqn:CG4a})-(\ref{eqn:CG4b}), and noting that the LHS leads to  a RHS where the density has
the different form noted above, but that ${\bf v}_R$ is {\it invariant} under this change of observer, for
each observer agrees on the absolute velocity of each piece of matter with respect to the local quantum foam.  Under
the change of observers, from $O$ to $O^\prime$,  (\ref{eqn:Vsum}) gives
\begin{equation} D_{ij}({\bf r},t) \rightarrow D_{ij}({\bf r+V}t,t) \mbox{\ \ and \ \ } 
 \Omega_{ij}({\bf r},t) \rightarrow \Omega_{ij}({\bf r+V}t,t).
\end{equation}
Then for the total or Euler fluid derivative in (\ref{eqn:CG4a}) we have for observer $O^\prime$ 
\begin{eqnarray}\label{eqn:covariant}
\frac{d D_{ij}({\bf r+V}t,t)}{dt} &\equiv&
\frac{\partial D_{ij}({\bf r+V}t,t)}{\partial t}+({\bf v}_O({\bf r}+{\bf V}t,t)-{\bf V}).{\bf \nabla}D_{ij}({\bf
r+V}t,t),\nonumber\\ &=&
\left.\frac{\partial D_{ij}({\bf r+V}t^\prime,t)}{\partial t^\prime}\right|_{t^\prime\rightarrow t}+
\left.\frac{\partial D_{ij}({\bf r+V}t,t^{\prime\prime}))}{\partial
t^{\prime\prime}}\right|_{t^{\prime\prime}\rightarrow t} + \nonumber \\&&\mbox{\ \ \ \ \ \ \ \ \ \ \ \ \ \ \ \  }
({\bf v}_O({\bf r}+{\bf V}t,t)-{\bf V}).{\bf
\nabla}D_{ij}({\bf r+V}t,t),\nonumber\\ &=&
({\bf V}.\nabla)D_{ij}({\bf r+V}t,t)+
\left.\frac{\partial D_{ij}({\bf r+V}t,t^{\prime\prime}))}{\partial
t^{\prime\prime}}\right|_{t^{\prime\prime}\rightarrow t} + \nonumber \\&&\mbox{\ \ \ \ \ \ \ \ \ \ \ \ \ \ \ \  }
({\bf v}_O({\bf r}+{\bf V}t,t)-{\bf V}).{\bf \nabla}D_{ij}({\bf r+V}t,t),\nonumber\\
&=&
\left.\frac{\partial D_{ij}({\bf r+V}t,t^{\prime\prime}))}{\partial
t^{\prime\prime}}\right|_{t^{\prime\prime}\rightarrow t} + 
{\bf v}_O({\bf r}+{\bf V}t,t).{\bf \nabla}D_{ij}({\bf r+V}t,t),\nonumber\\
&=&\left.\frac{d D_{ij}({\bf r},t))}{dt}\right|_{\displaystyle{{\bf r}\rightarrow {\bf r}+{\bf V}t}}
\end{eqnarray} 
as there is a key cancellation of two terms in (\ref{eqn:covariant}).  Clearly then all the terms on the LHS of 
(\ref{eqn:CG4a})-(\ref{eqn:CG4b}) have the same  transformation property. Then, finally, from the
form of the LHS,   both equations give the density dependent RHS, but which now  involves  the form
$  \left.\rho_O({\bf r},t)\right|_{\displaystyle{{\bf r}\rightarrow {\bf r}+{\bf V}t}}\mbox{\  }$, and this is
simply $\rho_{O^\prime}({\bf r},t)$ given above. 
If the observers coordinate axes do not have the same orientation then a time-independent orthogonal similarity
transformation $D \rightarrow SDS^T$,  $\Omega \rightarrow S\Omega S^T$, and ${ v}^i_R\rightarrow 
\sum_jS_{ij}v^j_R$ arises as well. Hence the description of the flow dynamics for observers in uniform relative
motion is Galilean covariant. While this transformation rule for the Euler derivative is not a new result, there
are some subtleties in the analysis, as seen above.  The subtlety arises because the change of coordinate
variables  necessarily introduces a time dependence in the observer descriptions, even if the flow is
inherently stationary.  

Finally, using an analogous argument to that in (\ref{eqn:covariant}),   we see explicitly that
the acceleration in (\ref{eqn:newaccel}) is also Galilean covariant under the above change of observer with transformation
(\ref{eqn:Vsum}), and indeed each of the three terms is separately covariant, with again the time derivative part of  the
middle term playing a key role, and then
${\bf g}({\bf r},t)\rightarrow{\bf g}({\bf r}+{\bf V}t,t)$ (in the case of observer axes with the same orientation). This
simply asserts that all observers actually agree on the gravitational  acceleration, up to the indicated trivial
translation effect caused by the motion of the observer. 

We now come to item (ii)  above, namely the more subtle but experimentally significant {\it approximate}
velocity superposition effect.  This approximate effect relates to the change in the form of the solutions of
(\ref{eqn:CG4a})-(\ref{eqn:CG4b}) when the matter density is in motion, as a whole, with respect
to the quantum-foam substratum, as compared to the solutions when the matter is, as a whole, at rest.   Already
even these descriptions involve a subtlety. Consider the case when a star, say, is `at rest' with respect to
the substratum. Then the flow dynamics in (\ref{eqn:CG4a})-(\ref{eqn:CG4b}) will lead to a position
and time dependent flow solution ${\bf v}({\bf r},t)$. But that flow leads to a position
and time dependent  ${\bf v}_R({\bf r},t)={\bf v}_0({\bf r},t)-{\bf v}({\bf r},t)$ on the RHS of
(\ref{eqn:CG4a})-(\ref{eqn:CG4b}), where  ${\bf v}_0({\bf r},t)$  is the velocity of the matter at
position ${\bf r}$ and time $t$ according to some specific observer's frame of reference\footnote{Here the
subscript is $0$ and not an $O$.  ${\bf v}_R({\bf r},t)$ was defined in (\ref{eqn:CG}). For matter described by a
density distribution it is appropriate to introduce the field ${\bf v}_0({\bf r},t)$. }. Hence the description of
the matter being `at rest' or `in motion' relative to the substratum  is far from simple.  In general, with
time-dependent flows, none of the matter will ever be `at rest' with respect to the substratum, and this
description is covariant under a change of observer.   In the case of a well isolated star existing in a
non-turbulent substratum we could give the terms `at rest as a whole' or `moving as whole' a well defined meaning
by deciding how the star as a whole, considered as a rigid body,  was moving relative to the more distant
unperturbed substratum.  Despite these complexities the solutions of
(\ref{eqn:CG4a})-(\ref{eqn:CG4b}) have, under certain special conditions, an approximate dynamical velocity
superposition effect, and these conditions actually occur for the earth, and have played a key role in observations
of absolute motion.  To see this effect we need to make some approximations in considering the form of the solutions
of (\ref{eqn:CG4a})-(\ref{eqn:CG4b}). First we note that  the vorticity  from (\ref{eqn:CG4b})  is 
small,  $\nabla \times {\bf v}\approx {\bf 0}$, as it is a `relativistic effect'. So for simplicity we shall assume
zero vorticity, and also neglect on the RHS the $({\bf v}_R/c)^2$ terms. We may then write ${\bf
v}=\nabla u$, and then  (\ref{eqn:CG4a}) reduces to   
\begin{equation}
\frac{\partial  u}{\partial t}=-\frac{1}{2}(\nabla u)^2-\Phi-\Phi_{DM},
\label{eqn:Cu}\end{equation}
where $\Phi$ is the Newtonian gravitational potential, as in (\ref{eqn:NGPhi}), and $\Phi_{DM}$ is an
effective `gravitational potential' that describes the dynamical `dark matter' effect,
\begin{equation}
\nabla^2\Phi_{DM}({\bf r},t)=4\pi G\rho_{DM}({\bf r},t),
\label{eqn:DMPhi}\end{equation}
with $\rho_{DM}$ defined in (\ref{eqn:DMdensity}), and so  $\Phi_{DM}[{\bf v}]$ depends 
functionally on $\nabla u({\bf
r},t)$. Of course  $\Phi$, like $\Phi_{DM}$, has the  form
\begin{equation}
\Phi({\bf r},t)=-G\int d^3 r^\prime\frac{\rho({\bf r}^\prime,t)}{|{\bf r}-{\bf
r}^\prime|}.
\label{eqn:ExplicitPhi}\end{equation} 
Eqn.(\ref{eqn:Cu}) is then an integro-differential equation determining the time evolution of $u({\bf r},t)$
from any given initial flow state $u({\bf r},t_0)$.  The $\Phi_{DM}$ term is an important non-Newtonian dynamical
feature of gravity, and leads to, for example, the bore hole $g$ anomaly,  the phenomenon of black holes, and the
non-Keplerian rotation of spiral galaxies.  

Eqn.(\ref{eqn:Cu}) gives for the time-evolution of the velocity field
\begin{equation}
{\bf v}({\bf r},t)={\bf v}({\bf r})-\nabla \int_0^t dt^\prime \left(\frac{1}{2}|{\bf v}({\bf
r},t^\prime)|^2+\Phi+\Phi_{DM}[{\bf v}]\right),
\label{eqn:veqn}\end{equation}
where clearly ${\bf v}({\bf r})$ is that flow at $t=0$.

The flow fields  have wavelike substructure. To see this suppose that (\ref{eqn:Cu}) has  a time
evolution
$u_0({\bf r},t)$ with corresponding velocity field
${\bf v}_0({\bf r},t)$.  Then we look for time-dependent perturbative solutions of (\ref{eqn:Cu})
with
$u=u_0+ \overline{u}$. To first order in $\overline{u}$ we then have
\begin{equation}
\frac{\partial {\overline u({\bf r},t)}}{\partial t}=-{\bf \nabla}\overline{u}({\bf r},t).{\bf
\nabla}u_0({\bf r},t).
\label{eqn:ueqn3}\end{equation}
This equation  has wave solutions of the form $\overline{u}({\bf r},t)=A\cos({\bf k}.{\bf
r}-\omega t+\phi)$ where $\omega({\bf k},{\bf r},t)={\bf v}_0({\bf r},t).{\bf k}$, for wavelengths and time-scales 
short compared to  the scale of changes in  ${\bf v}_0({\bf r},t)$. The local  phase velocity of these waves is then
${\bf v}_\phi={\bf v}_0$, and the local group velocity is ${\bf v}_g={\bf \nabla}_k\omega={\bf 
v}_0$. Then the velocity field is 
\begin{equation}
{\bf v}({\bf r},t)={\bf v}_0({\bf r})-A{\bf k}\sin({\bf k}.{\bf r}-w({\bf k},{\bf r},t)t+\phi).
\end{equation}
In general we have, perturbatively, the superposition of such waves, giving
\begin{equation}
{\bf v}({\bf r},t)={\bf v}_0({\bf r},t)-\int d^3k
A({\bf k}){\bf k}\sin({\bf k}.{\bf r}-w({\bf k},{\bf
r},t)t+\phi({\bf k})).
\end{equation}
 This perturbative analysis then suggests waves within waves, and with these waves interacting according to the
non-linear terms neglected in (\ref{eqn:ueqn3}), that is a turbulent fractal structure, where the equipotential
surfaces for $u$ have dimples upon dimples etc. Such wave effects have been detected \cite{NovaDM, RGC}.

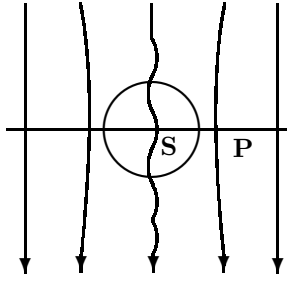
\begin{figure}[t]
\vspace{19mm}
\hspace{23mm}\setlength{\unitlength}{2.4mm}
\begin{picture}(20,10)
\thicklines
\put(13,17){\vector(0,-1){15}}
\put(27,17){\vector(0,-1){15}}
\put(20,10){\circle{5}}
\qbezier(16,17)(17,12)(16,2.5)
\put(16.1,3.1){\vector(0,-1){1}}
\qbezier(24,17)(23,12)(24,2.5)
\put(24,3.1){\vector(0,-1){1}}
\put(20,17){\line(0,-1){2}}
\qbezier(20,15)(20.5,14)(20,13)
\qbezier(20,13)(19.5,12)(20,11)
\qbezier(20,11)(20.5,10)(20,9)
\qbezier(20,9)(19.5,8)(20,7)
\qbezier(20,7)(20.5,6)(20,5)
\qbezier(20,5)(20.5,4)(20,3)
\put(20.1,3.1){\vector(0,-1){1}}
\put(20.5,8.6){${\bf S}$}
\put(24.5,8.5){${\bf P}$}
\put(12,10.0){\line(1,0){15.5}}
\end{picture}
\vspace{-4mm}
\caption{{\small  Velocity field ${\bf v}$, with asymptotic flow ${\bf V}$, expected from
(\ref{eqn:veqn}) showing greatest turbulence  effects along the direction parallel to  ${\bf V}$ and
through the bulk of the sun $S$.  Flow described by network of observers co-moving with the sun. On and near the plane
$P$, with normal
${\bf V}$, we have
${\bf v}_{in}.{\bf V}\approx {\bf 0}$.  The direction of absolute motion ${\bf V}$  of the solar system is
such that $P$ is very accurately the plane of the ecliptic. }}
\label{fig:turb}\end{figure}

Let us first consider the time evolution from  (\ref{eqn:veqn}) for the case of the sun undergoing an
absolute linear motion with absolute velocity $-\!{\bf V}$ (with respect to the substratum). This motion would be a
consequence of  galactic in-flows and the galactic orbital velocity of the solar system.  We shall neglect here any
time-dependence or inhomogeneity in ${\bf V}$, as we are interested here in the local effects caused by  the absolute
motion of the sun  through space. Let us start (\ref{eqn:veqn}) with
\begin{equation}
{\bf v}({\bf r}) = {\bf v}_{in}({\bf r}) +{\bf V},
\label{eqn:firstsum}\end{equation} 
where  
${\bf v}_{in}({\bf r})$ is a radial in-flow   which is an exact time-independent solution of (\ref{eqn:veqn})
when
${\bf V}={\bf 0}$, which exists if the matter density of the sun is taken to be spherically symmetric
\cite{NovaDM}. This 
${\bf v}({\bf r})$ has the asymptotic limit of
$+\!{\bf V}$, appropriate to the above absolute motion of the sun, and where in (\ref{eqn:firstsum}) we are using a 
network  of  observers co-moving with the sun. We can easily see how this absolute motion of the sun
affects the flow.  For a small time interval the change in
${\bf v}({\bf r},t)$ from (\ref{eqn:veqn}) is
\begin{equation}
\Delta{\bf v}({\bf r},t)=-\nabla({\bf v}_{in}({\bf r}).{\bf V})\Delta t+....
\label{eqn:vchange}\end{equation} 
This gives a growing change in ${\bf v}({\bf r},t)$, which to a first approximation is a non-uniform displacement of
the in-flow.   This is smallest in those regions where
${\bf v}_{in}.{\bf V}\approx {\bf 0}$, which is  near the plane
$P$ in   Fig.\ref{fig:turb}. The flow is thus expected to be most affected  along the direction parallel to 
${\bf V}$ and through the bulk of the sun $S$. The change in  (\ref{eqn:vchange})  cannot continue
indefinitely, and a better {\it ansatz} is to begin with a displaced in-flow as in 
\begin{equation}
{\bf v}({\bf r}) = {\bf v}_{in}({\bf r-a}) +{\bf V},
\label{eqn:vdisplaced}\end{equation}
where ${\bf a}$ parametrises  a uniform displacement to be estimated from the dynamics. This corresponds to the notion
that the in-flow is somewhat `dragged' or displaced by the absolute motion\footnote{This is completely different to the
old idea by Stokes of `entrainment', wherein the flow is supposed to have no ${\bf V}$ component in and near the earth.
This outdated idea arose from the erroneous conclusion that the Michelson-Morley 1887 experiment had failed to detect
absolute motion.}.  Using (\ref{eqn:vdisplaced}) in  (\ref{eqn:veqn}) gives, exactly, 
\begin{equation}
\frac{\partial {\bf v}}{\partial t}=\nabla\left(-{\bf v}_{in}({\bf r-a}).{\bf V}
+\Phi({\bf r-a})-\Phi({\bf r})+\Phi_{DM}({\bf r-a})-\Phi_{DM}({\bf r})\right),
\label{eqn:vrate}\end{equation}
where we have used the equation satisfied by the displaced in-flow  \hspace{1mm} ${\bf v}_{in}({\bf r-a})$. \hspace{3mm}
Then \newline the displacement
${\bf a}$ is to be determined by demanding that the time and spatial average  
\newline $<\!\!\frac{\partial {\bf v}}{\partial t}\!\!>_{t,{\bf r}}$ is  minimised.  Then the time-dependence is
reduced to that only of the necessary turbulence induced by the absolute motion of the matter through space.  Starting
the time-evolution with the flow in (\ref{eqn:firstsum}) will result in a relaxation to something like the flow
in (\ref{eqn:vdisplaced}) accompanied by excessive turbulence initially. To do better than (\ref{eqn:vdisplaced})
will require numerical modelling.  The approximate flow in (\ref{eqn:vdisplaced}) has an important property,
namely that the so-called `dark-matter' density is unchanged by a non-zero ${\bf V}$, except for the
displacement  $\rho_{DM}({\bf r})\rightarrow \rho_{DM}({\bf r-a})$.  As well in the limit
$\Phi_{DM}\rightarrow 0$ this flow gives exactly the same ${\bf g}$, up  to the translation effect, because of the
time-derivative term in (\ref{eqn:CG2}), as when  the sun is not in absolute motion, for the reasons discussed
above.

 Hence  the time-averaged flow is approximately ${\bf v}({\bf r}) = {\bf v}_{in}({\bf r}) +{\bf V}$, where well away 
from the sun  we can ignore any displacement  effect, with measure ${\bf a}$. This is even more accurate in the plane
$P$. This is the  dynamical superposition effect.   Now for the solar system, and thus the sun, the observed direction of
${\bf V}$  is such that $P$ is the plane of the ecliptic\footnote{That the direction of absolute motion of the solar
system is almost exactly normal to the plane of the ecliptic was discovered by Miller \cite{Miller}.  This is probably
not a coincidence as only then is the relativistic acceleration term in (\ref{eqn:newaccel})  a minimum. }.

\begin{figure}[t]
\vspace{17mm}
\hspace{52mm}\setlength{\unitlength}{0.8mm}
\begin{picture}(0,15)
\thicklines
\put(25,10){\vector(-1,0){15}}
\put(25,10){\vector(0,-1){29.5}}
\put(25,10){\vector(-1,2){15}}
\qbezier(0,0)(25,20)(50,0)
\put(30,-15){\Large $\bf v_{in}$}
\put(-17,10){\Large ${\bf v}_{tangent}$}
\put(17,30){\Large ${\bf v}_{N}$}
\end{picture}
\vspace{15mm} 
\caption{\small  Orbit of earth about the sun with tangential orbital velocity
${\bf v}_{tangent}$ and quantum-foam in-flow velocity  ${\bf v}_{in}$. Then ${\bf v}_{N}={\bf
v}_{tangent}-{\bf v}_{in}$ is the velocity of the earth relative to the quantum foam, after subtracting
the solar system cosmic velocity ${\bf V}$. }
\label{fig:orbit}\end{figure}
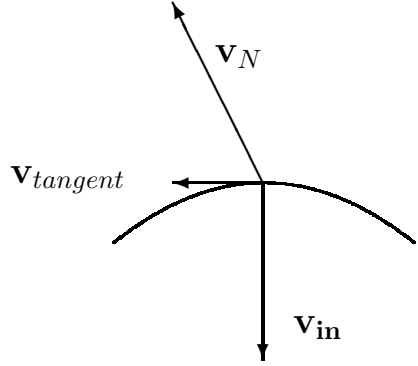

For the earth we may in the first instance ignore the mass of the earth, and treat it as a test particle
in motion through the above superposed flow determined  by the flow into the sun and the absolute linear motion
of the sun. Then for observers co-moving with the earth the
observed velocity is the vector sum of  the cosmic velocity of the
solar system, the in-flow of space past the earth into the sun, and the orbital velocity of the earth about the sun
(which enters with a minus sign for a co-moving observer). 
 Then, as in Fig.\ref{fig:orbit},
\begin{equation}\label{eqn:earthv}
{\bf v} \approx {\bf V} +{\bf v}_{in} -{\bf v }_{tangent}.
\end{equation}  
This neglects the flow component ${\bf v}_E$ caused by the matter of the earth. In the absence of absolute motion of the
earth this has a value of $11$km/s near the surface, and so is much smaller than the observed speed of absolute
motion of the earth.  To include at first approximation  ${\bf v}_E$  we can again use the displacement
{\it ansatz}, namely   ${\bf v}_E({\bf r}) \rightarrow {\bf v}_E({\bf r-b})$, where here ${\bf b}$ is the displacement
vector for the earth in-flow. Then (\ref{eqn:earthv}) becomes
\begin{equation}\label{eqn:earthv2}
{\bf v}({\bf r}) \approx {\bf V} +{\bf v}_{in} -{\bf v }_{tangent}-{\bf v }_E({\bf r-b}).
\end{equation}

For the earth this means also that  to this degree of approximation  the earth's absolute motion does not
affect the magnitude of the `dark-matter' effect within the earth,  causing only a displacement. This is 
important as in \cite{NovaDM} the  effects of absolute motion of the earth were neglected in analysing the  
bore-hole $g$ anomaly
data,  from which the parameter $\alpha$ was found  to be equal to the value of the fine structure constant, 
to within errors.
That analysis thus effectively assumed that the displacement effect was sufficiently small.
   
 The velocity superposition effect in (\ref{eqn:earthv}) was assumed in \cite{AMGE}, but  it was also assumed
implicitly by Miller \cite{Miller} in the analysis of his data, but there Miller did not include the ${\bf v}_{in}$
component as Miller was of course unaware of the flow theory of gravity.  For that reason a re-analysis of the Miller
scaling argument was required in \cite{AMGE}, and only then did the corrected Miller's scaling argument results for
the cosmic velocity of the solar system come into agreement with the new  velocity from analysis of the Miller data
using the refractive index effect.

 For circular orbits  of the earth about the sun  
$v_{tangent}$  and  $v_{in}$ are given by  
\begin{eqnarray}\label{eqn:QG7}
v_{tangent}&=&\sqrt{\displaystyle{\frac{GM}{R}}},\\  
v_{in}&=&\sqrt{\displaystyle{\frac{2GM}{R}}},\end{eqnarray}
while the net speed $v_N$ of the earth from the vector sum   ${\bf v}_N={\bf v}_{tangent}-{\bf
v}_{in}$  is 
\begin{equation}\label{eqn:QG7b}
v_{N}=\sqrt{\displaystyle{\frac{3GM}{R}}},\end{equation}     
where $M$ is the mass of the sun, $R$ is the distance of the earth from the
sun, and $G$ is Newton's gravitational constant.  The gravitational acceleration of the earth towards
the sun arises from  inhomogeneities in the $v_{in}$  flow component.  These expressions give
$v_{tangent}=30$km/s,  $v_{in}=42.4$km/s (at the earth distance) and
$v_{N}=52$km/s. As discussed in  \cite{AMGE} ${\bf v}_{in}$ is extractable from
Miller's 1925/26 air-mode Michelson interferometer experiment because Miller took data during four separate months of
the year, and over a year the vector sum of the three velocities varies. The extraction of $v_{in}$ from the
Miller data provided,  some 80 years after that most significant experiment,   the first experimental
confirmation of the new `in-flow' theory of gravity.

\section{ Conclusions\label{section:conclusions}}

The Newtonian theory of gravity and General Relativity have, perhaps somewhat surprisingly, only been tested within
very special circumstances; and when they failed in numerous other circumstances the experimental data was either
banned from the physics journals or  spurious explanations were invoked, the most infamous being of course
the `dark matter' explanation for the large non-Keplerian orbital velocities of the stars and gas clouds in the outer
regions of spiral galaxies.  The new theory of gravity has, however, not only agreed with the older two theories in
those special `successful' tests, but at the same time given explicit and checkable explanations and quantified
predictions for all the other `anomalies'.  In particular we have seen previously that the bore hole $g$ `anomaly'
data is directly linked to the spiral galaxy rotation `anomaly', and that this spatial self-interaction effect, as
it is now understood, may be studied in Cavendish laboratory experiments.   However the main result of this paper
has been to explain the dynamical effects behind the success of the velocity superposition effect, an effect
already assumed in the initial studies of the various absolute motion experiments \cite{CK,AMGE}. A feature of the new 
theory of gravity is that it also explains the Lense-Thirring `frame-dragging'  effect as a flow vorticity effect. 
In General Relativity the absolute rotation induced frame-dragging effect  has always been understood via the Mach
Principle: that motion of a particle is only meaningful when referred to the rest of the matter in the universe. In
the new theory of space and gravity we see that this is not so, namely that  motion is  observable with respect to a
substratum structure constituting the local space. The Gravity Probe B  satellite gyroscope experiment will thus explore
the vorticity effects, together with the geodetic effect, which is unrelated to vorticity.  For the vorticity effect
the new theory of gravity makes an additional prediction for the precession of the GP-B gyroscopes that is much larger
than that produced by the earth rotation-induced only vorticity predicted by General Relativity, which implies the
prediction that in its first experimental dynamical test GR will fail badly.   This is because in GR the effects of the very large, and already observed, absolute velocity of the earth
through the substratum of space is explicitly excluded: {\it absolute linear motion} is a concept absent from GR by
definition, and totally banned from physics.  Nevertheless the rotation induced
precessions are cumulative, while the linear motion induced precessions are not and have the periodicity of the orbit.
This may make the detection of the latter precessions difficult against a background of other effects having the same
periodicity. As well one must always emphasize that absolute linear motion is
completely consistent with the so-called `relativistic effects'; these are indeed caused by the dynamical effects of
absolute motion.   
 
We are entering an era where the full complexity of the phenomenon of gravitation will be, for the
first time, subjected to extensive experimental and observational  study, both by means of astronomical observations, for
example the spiral galaxy rotation data and also the globular cluster black hole mass observations, but also laboratory
experiments, such as those of the Cavendish kind where the effects associated with different matter shapes gives a
handle on the $\alpha$ dependent spatial self-interaction effect, which has actually plagued the accurate determination  of
$G$ for many decades, and resulted in $G$ being the most poorly measured fundamental constant. And also most significantly
we will see the beginning of systematic observations of the new gravitational wave phenomenon predicted by the new theory of
gravity, and  now apparent in existing experimental data. 

 As argued elsewhere the occurrence of
$\alpha$ as a second fundamental gravitational constant is a major development in our understanding of gravity, and is surely
indicating  that we are now entering the phenomena of quantum gravity, and that such effects are much larger than previously
predicted, that is, they do not manifest at the Planck scales. The argument that the Planck scales set the regime of
quantum gravity effects only followed when there was one dimensional gravitational constant, namely $G$.  But of course now we
have
$\alpha$  being the second but dimensionless gravitational constant.

\section{ References\label{section:references}}


\begin{thebibliography}{99}

\bibitem{NovaBook}  R.T. Cahill, {\it Process Physics: From Information Theory to Quantum Space and
Matter},  (Nova Science Pub., NY 2005), in book series {\it  Contemporary
Fundamental Physics},  edited by V.V. Dvoeglazov.

\bibitem{NovaDM} R.T. Cahill, {\it `Dark Matter' as a Quantum Foam In-Flow Effect}, in {\it Progress In Dark
Matter Research}, (Nova Science Pub., NY to be pub. 2004), physics/0405147;  R.T. Cahill, {\it Gravitation, 
the `Dark Matter' Effect and the Fine Structure Constant}, physics/0401047.

\bibitem{GQF}  R.T. Cahill, {\it Gravity as Quantum Foam In-Flow}, {\it Apeiron} {\bf 11}, No.1 (2004)1-52. 

\bibitem{RGC} R.T. Cahill, {\it Quantum Foam, Gravity and Gravitational Waves}, in {\it  Relativity,
Gravitation, Cosmology}, pp. 168-226, eds.  V. V. Dvoeglazov and A. A. Espinoza Garrido  (Nova Science Pub., NY
2004).

\bibitem{RC01}  R.T. Cahill,   {\it  Process  Physics: Inertia, Gravity and the Quantum},  {\it
Gen. Rel. and  Grav.} {\bf 34}(2002)1637-1656.

\bibitem{CK} R.T. Cahill  and  K. Kitto,  {\it  Michelson-Morley Experiments
Revisited and the Cosmic Background Radiation Preferred Frame},
{\it Apeiron} {\bf 10}, No.2 (2003)104-117. 

\bibitem{AMGE}   R.T. Cahill, {\it  Absolute Motion and Gravitational Effects}, {\it Apeiron} {\bf 11}, No.1 (2004)53-111.

\bibitem{Miller}  D.C.  Miller,     {\it Rev. Mod. Phys.} {\bf 5}(1933)203-242. 

\bibitem{GPB} R.T. Cahill, {\it  Novel Gravity Probe B Frame-Dragging Effect}, \newline physics/0406121.

\bibitem{GPBwaves} R.T. Cahill, {\it Novel  Gravity Probe B Gravitational Wave Detection}, \newline physics/0408097.

\bibitem{GPS}  R.T. Cahill, {\it Quantum-Foam In-Flow Theory of Gravity and the Global Positioning 
System (GPS)}, physics/0309016. 

\bibitem{Flandern} T. van Flandern, {\it Phys. Lett.} A{\bf 250}(1998)1-11.

\bibitem{Pugh} G. Pugh, in  {\it Nonlinear Gravitodynamics: The Lense
- Thirring Effect}, eds.   R. Ruffini and C. Sigismondi,
{\it World Scientific Publishing Company, 2003, pp 414-426}, (Based on a 1959 report).

\bibitem{Schiff}  L.I. Schiff, {\it Phys. Rev. Lett.} {\bf 4}(1960)215.

\bibitem{Ciufolini}  I. Ciufolini and E. Pavlis, {\it A Confirmation of the General
Relativistic Prediction of the Lense-Thirring Effect}, {\it Nature},
{\bf 431}(2004)958-960.

\bibitem{RK1} R.L.  Kirkwood,  {\it The Physical Basis of Gravitation}, {\it Phys. Rev.} {\bf
92}(6)(1953)1557.

\bibitem{RK2} R.L. Kirkwood,   {\it  Gravitational Field Equations}, {\it Phys. Rev.} {\bf
95}(4)(1954)1051.

\bibitem{CM} F. Cardone and R. Mignani, {\it Intl. J. Mod. Phys.}, {\bf 14}, 24 (1999)3799-3811.



\end{thebibliography}
\end{document}